\documentstyle[twocolumn,floats,aps,prb]{revtex}

\input epsf
\epsfverbosetrue

\begin{document}

\twocolumn[\hsize\textwidth\columnwidth\hsize\csname
@twocolumnfalse\endcsname

\title {\bf Theoretical examination of stress fields in Pb(Zr$_{0.5}$Ti$_{0.5}$)O$_3$}
\author{Nicholas J. Ramer,$^{1}$ E. J. Mele$^{2,3}$ and Andrew M. Rappe$^{1,3}$}
\address{$^1$Department of Chemistry, $^2$Department of Physics and \\$^3$Laboratory for Research on the Structure of Matter\\ University of Pennsylvania, Philadelphia, PA 19104}
\date{\today}
\maketitle

\begin{abstract} 
In this paper, we develop a rigorous formulation of the local stress field.  This approach can be used in conjunction with any first-principles method to study stress fields in complex bonded systems.  In particular we investigate the induced stress fields resulting from the homogeneous deformations of tetragonal PbTiO$_3$ and rhombohedral PbZrO$_3$.  As an extension of these findings we also compute the induced stress fields resulting from homogeneous deformation of the (100) and (111) orderings of Pb(Zr$_{0.5}$Ti$_{0.5}$)O$_3$.  The stress-field responses in these four materials are compared and their piezoelectric responses are discussed.\vspace{5 mm}

\vspace{5 mm}
\end{abstract} 
]
\newpage

\centerline{\bf I.  INTRODUCTION}\vspace {3 mm}

When mechanical stress is applied to a piezoelectric material, microscopic atomic rearrangements occur which give rise to a change in the macroscopic electric polarization of the material.  Conversely, application of a voltage across a piezoelectric material produces an internal strain within it.  In either case, it is the material's internal structural behavior (and therefore spontaneous polarization) under applied or induced stress that underlies the piezoelectric response.  

The ability of piezoelectric materials to interconvert electrical and mechanical energy lies at the foundation of many electro-optic and electro-acoustic devices.  The use of piezoelectric materials in these devices stems from the need to monitor the magnitude of induced or applied electrical response through the device.  One of the clearest examples of such an application is seen in the constant tunneling voltage mode of the scanning-tunneling microscope (STM).\cite{STM}  The STM scanning needle is attached to a cylinder of piezoelectric ceramic.  A voltage is applied across the gap between the needle tip and the scanned surface.  If there are any structural variations in the scanned surface, this will cause a change in the voltage.  Using a feedback loop, the piezoelectric ceramic is either compressed or expanded by altering the current passing through the ceramic so as to adjust the height of the scanning tip relative to the surface until the originally applied voltage is restored.  Monitoring the height of the tip provides information on the structure of the scanned surface.  Piezoelectric materials also play a vital role in electro-acoustic transducers.  In these devices the piezoelectric material act as an interpreter for the incoming (or outgoing) sound wave and the outgoing (or incoming) electric signal.  These types of devices have uses in underwater and medical ultrasonic imaging.

In this paper, we present a first-principles investigation of the distorted perovskite materials PbTiO$_3$ and PbZrO$_3$ at zero temperature and study the spatial variation of their stress-field responses to an externally applied uniform strain.  In addition, stress-field studies of the (100) and (111) orderings of the solid solution Pb(Zr$_{0.5}$Ti$_{0.5}$)O$_3$ (PZT) are also reported.  Our selection of PZT is motivated by the large quantity of experimental studies describing the existence of a strong piezoelectric response in various composite PZT ceramics.\cite{Jaffe}  In Section II we briefly describe the formalism for construction of the local stress fields.  In Section III we present results for the computation of local stress fields induced by a uniform uniaxial deformation.  A discussion of the local stress fields produced in the simple perovskite crystals and the more complex PZT superlattices is provided in Section IV.  We review the main conclusions of this study in Section V.\vspace {5 mm}
\centerline{\bf II. STRESS-FIELD FORMALISM}\vspace {3 mm}

It is central to the study of any piezoelectric crystal to understand the material's internal response to an externally applied strain.  Experimentally, a particular crystal's structural response to an applied strain can be measured using various diffraction techniques \cite{JonaShirane} or direct measurement of the changes in the dimension of the sample by electrical capacitance or optical interference.\cite{Ikeda}  In order to ascertain the effect strain has upon piezoelectric response, these techniques are paired with high-field measurement of strain hysteresis and polarization change.  More recently, field-induced strain has been measured using a displacement magnification technique.\cite{Saito}  However because the formation and testing of many of these strained materials is complicated and possibly destructive,\cite{Ikegamidiff} it is advantageous to have a concise theoretical method from which information concerning the microscopic response of a system to an external strain can be extracted.  

We consider the response of a system to a homogeneous long wavelength deformation (scaling transformations describing pure dilation, strain or shear).  For an interacting system of atoms, the introduction of any of these uniform deformations can induce a force distribution on all the structural degrees of freedom of the unit cell.  Within a harmonic theory, the induced atomic force distribution, $\vec{F}$, contains all the relevant information about the redistribution of the external stress within the cell.  Therefore, the starting point for the computation of the local stress fields is the calculation of the induced force distribution.  This can be accomplished using local density functional theory.  These theoretical methods have proven to be very successful for studying structural phenomena in a broad class of condensed phases.  Below we outline our method to compute the stress field given the local force density.  Once the stress field has been constructed, correlations between the elastic response and the structural features of the system can be made.  A more detailed explanation of the method will be presented elsewhere.\cite{Mele97}

We begin with the principle of virtual work in the presence of an induced force distribution $\vec{F}_m$.  Any set of displacements of the nuclear coordinates $\vec{u}$ for a particular interacting system produces a variation of the energy according to

\begin{eqnarray}
\delta U &=& \sum_{m=1}^{N_a} \vec{F}_m \cdot\vec{u}_m 
\end{eqnarray}
\noindent where $m$ represents the $m$-th ion of the interacting system.  It is useful to convert the displacements and forces to continuous fields:
\begin{eqnarray}
\delta U &=& \frac{1}{\Omega}\int_{\rm{cell}} d^3r\vec{F}\left(\vec{r}\,\right)\cdot\vec{u}\left(\vec{r}\,\right)\nonumber \\
 &=& \sum_{\vec{G}}\vec{F}\left(-\vec{G}\right)\cdot\vec{u}\left(\vec{G}\right)
\end{eqnarray}
In the last line we have used lattice translational symmetry to transform the force and displacement distributions to a reciprocal space representation.  The principle of virtual work can be recast in terms of the internal strains in the structure
\begin{eqnarray}
\delta U &=& \int_{\rm{cell}} d ^3r\stackrel{\Rightarrow}{\sigma}\left(\vec{r}\,\right)\cdot\stackrel{\Rightarrow}{\epsilon}\left(\vec{r}\,\right)\nonumber \\
 &=& \Omega\sum_{\vec{G}}\stackrel{\Rightarrow}{\sigma}\left(-\vec{G}\right)\cdot\stackrel{\Rightarrow}{\epsilon}\left(\vec{G}\right)
\end{eqnarray}
\noindent where $\stackrel{\Rightarrow}{\sigma}$ is the 6-component stress field tensor and $\stackrel{\Rightarrow}{\epsilon}$ is the 6-component strain tensor.  (Since only the contraction of two rank-2 tensors is required in equation (3), the tensors are represented as length-6 vectors for simplicity.)

The direct space components of the strain tensor can be directly related to the displacement field,

\begin{eqnarray}
\epsilon_{n}\left(\vec{r}\,\right) = \vec{\nabla}\cdot\Gamma_{n}\cdot\vec{u}\left(\vec{r}\,\right)
\end{eqnarray}

\noindent where $\epsilon_{n}$ is the $n$-th component of the strain tensor ($n=1,\ldots,6$) and $\Gamma_n$ is the 3$\times$3 matrix determining the symmetry of $\epsilon_{n}$.  Transforming this relationship into Fourier space gives the particularly convenient result

\begin{eqnarray}
\epsilon_{n}\left(\vec{G}\right) = \vec{G}\cdot\Gamma_{n}\cdot\vec{u}\left(\vec{G}\right)
\end{eqnarray}

Combining equation (5) for all 6 components of the strain tensor gives

\begin{eqnarray}
\stackrel{\Rightarrow}{\epsilon}\left(\vec{G}\right) = {\cal P} \cdot \vec{u}\left(\vec{G}\right)
\end{eqnarray}

\noindent or

\begin{eqnarray}
{\cal P}^{-1}\cdot\stackrel{\Rightarrow}{\epsilon}\left(\vec{G}\right) = \vec{u}\left(\vec{G}\right)
\end{eqnarray}

Inserting this relationship into the last line of equation (2) and equating lines (2) and (3) yields

\begin{eqnarray}
\delta U &=& \Omega\sum_{\vec{G}}\vec{F}\left(-\vec{G}\right)\cdot{\cal P}^{-1}\cdot\stackrel{\Rightarrow}{\epsilon}\left(\vec{G}\right)\\
&=&\Omega\sum_{\vec{G}}\stackrel{\Rightarrow}{\sigma}\left(-\vec{G}\right)\cdot\stackrel{\Rightarrow}{\epsilon}\left(\vec{G}\right)
\end{eqnarray}

Equating the arguments of the summations in equations (8) and (9) yields all 6 components of the local stress field in reciprocal space:

\begin{eqnarray}
\stackrel{\Rightarrow}{\sigma}\left(-\vec{G}\right)=\vec{F}\left(-\vec{G}\right)\cdot {\cal P}^{-1}
\end{eqnarray}

\noindent By transforming according to

\begin{eqnarray}
\stackrel{\Rightarrow}{\sigma}\left(\vec{r}\,\right) = \sum_{\vec{G}}\stackrel{\Rightarrow}{\sigma}\left(\vec{G}\right)e^{i\vec{G}\cdot\vec{r}}
\end{eqnarray}

\noindent we can construct the stress distribution in direct space, and this result can be used to generate a map of the spatial distribution of the $n$-th stress field of the system.

Vanderbilt\cite{VanderbiltPC} has correctly pointed out that this formalism only enables computation of the stress-field components which possess the periodicity of the unit cell.  In addition the $\vec{G}$=0 stress tensor can be computed by the approach of Nielsen and Martin.\cite{NielMart1,NielMart2}  Development of a method for the computation of the stress-field components which are uniform in one or two dimensions and of finite wavelength in the others is in progress.\vspace {5 mm}
\centerline{\bf III.  RESULTS}\vspace {3 mm}

The first-principles calculations presented in this paper are performed within density functional theory,\cite{DFT} and the local density approximation (LDA) is used to describe the electron-electron interactions.  For the solid-state calculations, the single electron wave functions are expanded in a plane-wave basis using a cutoff energy of 50 Ry.

To describe the electron-nuclear interaction, optimized pseudopotentials\cite{RappePS} in fully separable nonlocal form\cite{Kleinman} are used.  An additional feature of our nonlocal pseudopotentials\cite{RamerPS} is their improved transferability over a wide range of electronic configurations.  We have been able to exploit the flexibility contained in the separation of the local and non-local parts of the pseudopotential.  By designing the form of the local potential so that the pseudo-eigenvalues and all-electron eigenvalues agree at an additional charge state, it is possible to improve the transferability of the potential across the charge states lying between the original reference state and this second charge state.

Due to the need for high accuracy when examining ferroelectric phenomena, semi-core shells are included in the generation of the pseudopotentials.  We include as valence states the 3$s$ and 3$p$ for Ti and the 4$s$ and 4$p$ for Zr.  The 5$d$ shell is included for Pb.  Furthermore, scalar relativistic effects are included in the generation of the Pb pseudopotential.\cite{Koelling}  For each metal, a pseudopotential is constructed using a designed local potential with the addition of a square well within the core region.  By doing so, we are able to achieve excellent transferability of the pseudopotential over a variety of charge and excited states.  For each of the metals, excellent agreement of the pseudo-eigenvalues and total energy differences with the all-electron results is achieved for charge states of +4 to neutral.  The oxygen pseudopotential is constructed using the $s$ angular momentum channel as the local potential.  

Brillouin zone integrations for PbTiO$_3$ and PbZrO$_3$ were done using a $4\times4\times4$ Monkhorst-Pack $k$-point mesh.\cite{Monkhorst}  For tetragonal PbTiO$_3$ and rhombohedral PbZrO$_3$ this yields 6 and 10 $k$-points, respectively in the irreducible wedge of the Brillouin zone.  For the PZT crystals, the same $k$-point mesh was folded into the smaller Brillouin zone, resulting in 3 irreducible $k$-points for (100)-PZT, which was taken to have tetragonal symmetry, and 5 irreducible $k$-points for (111)-PZT, which was taken to have rhombohedral symmetry.  It should be noted that to compute the nonuniform force distribution resulting from the application of a uniform external stress, additional $k$-points were needed due to the broken symmetries in the distorted structure.

The calculations involving the rhombohedral PbZrO$_3$ and (111)-PZT deserve closer consideration.  Crystallographically, the zero-temperature form of PbZrO$_3$ is orthorhombic with 40 atoms per unit cell.\cite{FujishitaI,FujishitaII,Egami}  In order to simplify our comparisons of the local stress fields, the structure of PbZrO$_3$ was fully relaxed within the rhombohedral symmetry of zero-temperature ferroelectric phase of PbZrO$_3$ with small Ti doping.\cite{Jaffe}   Rhombohedral strains of the unit cell shape away from cubic were neglected since they have been shown to be quite small ($\sim$ 0.1$^\circ$).\cite{Whatmore}  In the case of (111)-PZT, rhombohedral strains were also neglected to simplify analysis of the induced local stress fields.

As part of the local stress-field calculations, complete structural relaxations of both internal coordinates and crystal lattice constants were completed for the PbTiO$_3$ and PbZrO$_3$ distorted perovskite structures.  We report our atomic positions and lattice constants for tetragonal PbTiO$_3$ and rhombohedral PbZrO$_3$ in Table I and compare these results with previous theoretical and experimental values where possible.  The absence of experimental values for the rhombohedral PbZrO$_3$ is explained by our simplification of the PbZrO$_3$ structure as described above.  In the case of PbTiO$_3$, our agreement to previously reported theoretical \cite{Garcia} and experimental results \cite{Friere} is quite good and is within the error expected from density functional solid-state calculations done within LDA.  For the rhombohedral PbZrO$_3$, our results agree quite well with the theoretical results of Singh.\cite{Singh}  (In Singh's work, the lattice constant of rhombohedral PbZrO$_3$ was not relaxed.  The theoretically determined lattice constant of the high temperature cubic perovskite was used instead.)

\begin{table}[t]
\caption{Computed and experimental equilibrium lattice constants and atomic positions for tetragonal PbTiO$_3$ and rhombohedral PbZrO$_3$.}

\begin{tabular}{lccc}
 &Present&Theory&Experiment\\ \hline
\vspace{0.05 in}\\
 \multicolumn{4}{c}{\bf{PbTiO$_3$}} \\
\vspace{0.05 in}\\
$a$$\left(\right.$\AA$\left.\right)$&3.870&3.862\tablenote[1]{Reference\protect[20].}&3.905\tablenote[2]{Reference\protect[21].}\\
$c/a$&1.063&1.054&1.063\\
$x\left(\right.$Ti$\left.\right)$&0.531&0.537&0.540\\
$x\left(\right.$O$_{1}$,O$_{2}$$\left.\right)$&0.604&0.611&0.612\\
$x\left(\right.$O$_{3}$$\left.\right)$&0.098&0.100&0.112\\
\vspace{0.05 in}\\
\multicolumn{4}{c}{\bf{PbZrO$_3$}} \\
\vspace{0.05 in}\\
$a$$\left(\right.$\AA$\left.\right)$&4.143&4.12\tablenote[3]{Reference\protect[22].}& \\
$x\left(\right.$Zr$\left.\right)$&0.540&0.545& \\
$x\left(\right.$O$_{1}$$\left.\right)$&0.057&0.061& \\
$y,z\left(\right.$O$_{1}$$\left.\right)$&0.583&0.590& \\
 & & & \\
\end{tabular}
\end{table}

Atomic and lattice relaxations were also performed for the PZT superlattices.  The experimental lattice constants as well as theoretical lattice constants and relaxed atomic positions are contained in Table II.  Experimental values for the (100)-PZT ceramic are taken for  the tetragonal 50-50 batch composition PZT ceramic according to Jaffe $et\,\,al.$ (ceramic {\bf 3} using the notation from reference [2]).  The experimental values for the (111)-PZT ceramic are taken for a rhombohedral PZT ceramic close to the 50-50 batch composition (ceramic {\bf 5} using the notation from reference [2]).

\begin{table}[t]
\caption{Computed equilibrium lattice constants and atomic positions for (100)-PZT and (111)-PZT. Experimental lattice constants are given for randomly ordered PZT ceramics close to the 50-50 batch composition.  See text for description.}
\begin{tabular}{lrc}
 &\multicolumn{1}{c}{Present}& Experiment \\ \hline
\vspace{0.05 in}\\
\multicolumn{3}{c}{\bf{(100)-Pb(Zr$_{0.5}$Ti$_{0.5}$)O$_3$}}\\
\vspace{0.05 in}\\
$a$$\left(\right.$\AA$\left.\right)$&8.313&8.279\tablenote[1]{Reference\protect[2].} \\
$c/a$&0.480&0.487\\
$x\left(\right.$Pb$_{2}\left.\right)$&0.468& \\
$x\left(\right.$Ti$\left.\right)$&0.211& \\
$x\left(\right.$Zr$\left.\right)$&0.714& \\
$x\left(\right.$O$_{1}$$\left.\right)$&-0.049& \\
$x\left(\right.$O$_{2}$,O$_{3}$$\left.\right)$&0.185& \\
$x\left(\right.$O$_{4}$$\left.\right)$&0.424& \\
$x\left(\right.$O$_{5}$,O$_{6}$$\left.\right)$&0.660& \\
\vspace{0.05 in}\\
\multicolumn{3}{c}{\bf{(111)-Pb(Zr$_{0.5}$Ti$_{0.5}$)O$_3$}}\\
\vspace{0.05 in}\\
$a$$\left(\right.$\AA$\left.\right)$&8.043&8.164$^{\rm{a}}$\\
$x\left(\right.$Pb$_{2}$$\left.\right)$&0.498& \\
$x\left(\right.$Ti$\left.\right)$&0.237& \\
$x\left(\right.$Zr$\left.\right)$&0.737& \\
$x\left(\right.$O$_{1}$$\left.\right)$&-0.014& \\
$z\left(\right.$O$_{1}$$\left.\right)$&0.221& \\
$x\left(\right.$O$_{4}$$\left.\right)$&0.470& \\
$z\left(\right.$O$_{4}$$\left.\right)$&0.721& \\
 & & \\
\end{tabular}
\end{table}

We studied the induced local stress fields by calculating the internal force distribution induced by a uniform external strain.  The force distribution is obtained from first-principles density functional theory within the LDA and the Hellmann-Feynman theorem.\cite{Hellmann,Feynman}  Since our stress-field formalism relies on the fact that any deformation must not take the system beyond lowest order in gradients of the total energy, attention must be paid to the magnitude of the deformations.  We have found that $\pm$0.2\% deformations in lattice lengths and $\pm$0.5$^\circ$ in lattice angles are within the harmonic limit of the potential energy.  Extension of this work beyond harmonic order is a promising direction for the study of recently discovered piezoelectric single crystals which exhibit large reversible strains.\cite{Park}

There are 6 homogeneous deformations which can be made to any crystal: dilation having the symmetry of $x^2+y^2+z^2$, uniaxial strains---tetragonal with symmetry $2x^2-y^2-z^2$ and orthorhombic with $y^2-z^2$ symmetry, and elementary shear operations ($xy$, $xz$, and $yz$).  Based on crystal symmetry considerations, certain homogeneous deformations are degenerate and can easily be constructed from the other deformations.  For a crystal subjected to any of the 6 homogeneous deformations one finds a local $internal$ stress field in all 6 stress components.  It is important to note that for a given deformation, the induced stress fields corresponding to the 5 other deformations must each integrate to zero over the entire unit cell.  As an example, for a tetragonal deformation, there is, of course, a tetragonal local stress-field response.  In addition, one induces local dilation, orthorhombic and shear responses.  The latter 5 responses, due to symmetry constraints, must each integrate to zero.  

For brevity, we only report the local stress fields induced by an applied tetragonal uniaxial stress.  In particular we focus our discussion here on the internal dilation and tetragonal stress-field responses induced by a tetragonal compression.   A more comprehensive presentation of all the induced local stress fields originating from each of the 6 deformations is in progress.\cite{BigPaper}

To simplify the visualization of the stress fields, we have chosen to show only the regions of highest induced local stress.  For the dilation response to a uniaxial compressive deformation, light blue isosurfaces correspond to regions in the unit cell undergoing compression in all directions and  pink isosurfaces correspond to expansion in all directions.  For the tetragonal response to a tetragonal compressive deformation, light blue regions correspond to prolate response (expansive along the axial direction but compressive along the equatorial directions) and pink regions correspond to oblate response (compressive along the axial direction but expansive along the equatorial directions).\vspace {3 mm}

\centerline{\bf A. Tetragonal PbTiO$_3$}\vspace {1 mm}
Figures IA and IB show the dilation and tetragonal local stress fields, respectively, produced in response to a uniaxial tetragonal deformation of the zero-temperature equilibrium structure of tetragonal PbTiO$_3$.  The (100) lattice direction contains the Pb atoms at the lower left forward corner and upper left forward corners of the unit cell.  It is instructive to recognize the regions under the highest induced stress for both stress components lie along the (100) direction, as this coincides with the direction of ferroelectric distortion in the relaxed crystal and the direction of applied stress.

{\bf Dilation Response:} The most salient feature of this response is its location relative to the Pb and Ti atoms.  We find the region of highest induced stress does not involve the Pb atoms but instead straddles the Ti atom and is oriented along the (100) direction.  The spatial representation of this component of the local stress field indicates that under uniaxial tetragonal compression, the response of the crystal would be to compress the volume in the half of the TiO$_6$ octahedron lying above the Ti atom.  A commensurate expansion of the volume in the half lying below the Ti atom is also seen in this response.  The combination of these volume deformations would result in the shifting of the Ti atom toward the upper half of the octahedron.  In contrast, the oxygens lying collinear with the regions of highest induced stress would shift in the ($\overline{1}$00) direction, opposing the motion of the Ti atom.  This oxygen motion is also indicative of an expansion of the lower half of the TiO$_6$ octahedron and a compression of the upper half.  Finally, there is no significant stress-field response involving the oxygens lying equatorial to the Ti atom.

{\bf Tetragonal Response:} Once again we find the region of highest induced stress does not involve the Pb atoms but surrounds the Ti atom and lies along the (100) direction.  Although the interpretation of this stress component is different from the dilation response, its net atomic response is similar.  In the upper half of the TiO$_6$ octahedron lying along the (100) direction, there is an oblate response which would shift the Ti along the (100) direction toward the upper half of the octahedron.  In addition to the oblate response, there is a prolate response in the lower half of the TiO$_6$ octahedron.  The effect of these volume deformations would be to elongate the lower half of the oxygen octahedron while shortening the upper half.  The resulting oxygen motion is consistent with the volume deformations.\vspace {3 mm}

\centerline{\bf B. Rhombohedral PbZrO$_3$}\vspace {1 mm}
Figures IIA and IIB show the dilation and tetragonal local stress fields produced in response to a uniaxial tetragonal deformation of the zero-temperature equilibrium structure of rhombohedral PbZrO$_3$.  The (111) atomic distortion direction contains the Pb atoms at the lower left forward and upper right rear corners of the unit cell.  It is useful to recognize that some regions under the highest induced stress for both stress components lie not exclusively along the (111) distortion direction, but also along the (100) direction, the direction of uniaxial strain.

{\bf Dilation Response:} In this stress field, as in the dilation field for PbTiO$_{3}$, there is pairing of compressive and expansive regions surrounding the central metal atom.  However, there is a clear difference between the dilation response for the 2 crystals.  In PbZrO$_3$ response we find a compression of the volume above the Zr and an expansion of the volume below the Zr atom.  These regions of the stress fields are some combination of the (111) ferroelectric distortion direction and (100) uniaxial strain direction.  The corresponding motion of the Zr is toward the upper half of the ZrO$_6$ octahedron, along a direction lying between the (111) and (100) directions.  In addition to the motion of the Zr atom, the O atoms lying axial to the Zr atom are moving in a direction opposing the Zr atom motion.  This is clearly seen in the combination of the compressive and expansive deformations of the volumes within the unit cell. 

{\bf Tetragonal Response:}  This response has some features in common with the dilation response of the PbZrO$_{3}$ crystal described above.  Here again there is a pairing of regions oriented along a combination of the (111) and the (100) directions.  These regions are also centered at the Zr atom.  Specifically, we find a prolate region in the lower half of the ZrO$_6$ octahedron and an oblate region in the upper half of the octahedron.  The resulting motion of the O atoms lying axial to the Zr atoms is similar to the the motion of the O atom in the dilation response of this crystal.\vspace {3 mm}  

\centerline{\bf C. (100)-Pb(Zr$_{0.5}$Ti$_{0.5}$)O$_3$}\vspace{1 mm}
Figures IIIA and IIIB show the dilation and tetragonal local stress fields produced in response to a uniaxial tetragonal deformation of the zero-temperature equilibrium structure of (100)-Pb(Zr$_{0.5}$Ti$_{0.5}$)O$_3$.  The (100) lattice direction contains the Pb atoms at the lower left forward corner and upper left forward corners of the unit cell.  (Our restriction to tetragonal symmetry forces the ferroelectric distortion to be along (100).)  In both reported responses, the highest induced stress field lies along the (100) direction parallel to the ferroelectric distortion direction and the (100) direction of uniaxial strain.

{\bf Dilation Response:} For this super-cell, an alternating pattern of compressive and expansive responses is found lying along the 4-fold rotation axis.  In this response to the tetragonal uniaxial compression, the highest induced stress fields are reminiscent of the stress field in the dilation responses of both distorted perovskite structures.   However, while the response in the PbTiO$_3$ sub-unit is remarkably similar to the dilation response of the tetragonal PbTiO$_3$ described above, the response in the PbZrO$_3$ half of the unit cell shows an important difference from the dilation response of the rhombohedral PbZrO$_3$.  In the simple distorted zirconate, Zr atomic motions along a composite direction lying between the (100) and (111) directions became active under tetragonal deformation.  As a result of our imposition of tetragonal symmetry, we find response in the (100)-PZT superlattice lying exclusively along the (100) direction indicating atomic motions purely in the (100) direction.  In each oxygen octahedron, there is a compression of the lower half along with an expansion of the upper half.  This leads to Ti and Zr motions toward the lower halves of the octahedra.  The axial O atoms show opposing motion to the Ti and Zr atoms.

{\bf Tetragonal Response:} In this response to the uniaxial deformation, the induced stress field shows an alternating pattern of prolate and oblate components lying along the 4-fold rotation axis.  The resulting motion of the atoms is consistent with the dilation response.\vspace {3 mm}

\centerline{\bf D. (111)-Pb(Zr$_{0.5}$Ti$_{0.5}$)O$_3$}\vspace{1 mm}
Figures IVA and IVB show the dilation and tetragonal local stress fields produced in response to a uniaxial tetragonal deformation of the zero-temperature equilibrium structure of (111)-Pb(Zr$_{0.5}$Ti$_{0.5}$)O$_3$.  The (111) atomic distortion direction contains the Pb atoms at the lower left forward and upper right rear corners of the unit cell.  The (100) lattice direction contains the Pb atom at the lower left forward and upper left forward Pb atoms in the unit cell.  There are 2 crystallographically unique Pb atoms lying along the (111) direction in addition to the Ti and Zr atoms.  Due to ferroelectric distortion, the Ti atom has shifted along ($\overline{111}$) toward the Pb atom at the lower left forward corner and away from the Pb in the center of the figure.  The Zr atom has moved toward the central Pb atom along the ($\overline{111}$).

{\bf Dilation Response:} Each individual region of highest induced stress field in the response is oriented perpendicular to the (111) atomic distortion direction.  Unlike the responses for the previously mentioned crystals, we find the Ti and Zr atoms of (111)-PZT have negligible involvement.  Instead we report alternating expansion and compression regions consisting of entire Pb$_2$O$_4$ octahedra.  These octahedra are comprised of 2 adjacent Pb atoms lying along the (100) lattice direction and 4 equatorial O atoms.  The overall pattern of the expansions and compressions shows the (111) stacking of the entire superlattice.  Focusing on the Pb at the lower left forward corner, we find expansion of the Pb$_2$O$_4$ octahedron above this atom and contraction of the Pb$_2$O$_4$ octahedron below, resulting in largely downward motion of this Pb atom.  Analogous reasoning shows that the dilation stress field causes upward motion of the central Pb atom.  The expansion and contraction of the Pb$_2$O$_4$ octahedra also gives rise to equatorial oxygen atomic motion in the (100) plane.  The O atoms not involved in the Pb$_2$O$_4$ octahedra also exhibit atomic motion.  Due to the ellipsoidal shape of the stress-field regions, the O atoms lying beneath the Ti atoms are moving primarily upward along the (100) while the O atoms lying beneath the Zr atoms are moving mostly downward.

{\bf Tetragonal Response:} In this response we report a feature of the stress field not found in any of the other responses for the other materials.  Namely, we find prolate regions in the titanate sub-units alternating with oblate regions in the zirconate sub-units.  In particular, the responsive regions encompass the entire MO$_6$ octahedra of their respective sub-units.  In each Zr sub-unit, there is a prolate response while in the TiO$_6$ octahedron there is an oblate response.

\vspace {5 mm}
\centerline{\bf IV.  DISCUSSIONS}\vspace {3 mm}

As a general guideline for analysis, it is important to understand the relationship between the ferroelectric distortion for a particular material and the direction of uniaxial stress.  It is the combination of these effects that will either enhance or diminish the effect of the ferroelectric distortion and therefore the extent of piezoelectric response.  For all of the above reported responses the direction of uniaxial stress is (100).  

As previously stated, the ferroelectric distortion for PbTiO$_3$ lies along the (100).  Although pure PbTiO$_3$ is not piezoelectric, piezoelectric response has been reported for a PbTiO$_3$ ceramic doped with 2.5 mol\% LaO$_{0.5}$ and 1.0 mol\% MnO$_2$.\cite{Ikeda}  While the piezoelectric response of this material is smaller than that of the PZT solid solution ceramics, it is interesting to note that the calculated stress fields for the undoped material show the ability of the crystal to exhibit such a response upon application of a uniaxial strain.  By examination of the results in Table I, it is clear that the relative positions of the Ti and O atoms underlie the ferroelectric response in this crystal.  In particular, we find the Ti atom lies $\sim$0.28\AA$\:$ below the equatorial oxygen plane.  This means that in the zero-temperature tetragonal phase the volume beneath the Ti atom is compressed while the volume above the Ti atom is expanded with respect to the cubic prototype structure.  We have found in both reported responses a tendency for the volume beneath the Ti atom to expand and the volume above the Ti atom to compress upon application of a uniaxial compression.  These volume deformations are indicative of a structural phase transformation from tetragonal to a para-electric cubic phase.  Since the cubic structure shows no ferroelectric response, the application of a strain that returns the tetragonal PbTiO$_3$ toward the cubic phase will reduce the ferroelectric properties and therefore underlies the piezoelectric response of this material upon uniaxial compression.

The structure of rhombohedral PbZrO$_3$ shows an atomic distortion along the (111) direction.  Upon application of the tetragonal strain, we find the highest induced stress field lying along a combination of the strain direction and the atomic distortion direction.  It is clear in these responses that, for this crystal, the directions of uniaxial strain and atomic distortion are competitive influences on the atomic displacements.  Despite the complexity of the predicted atomic motions, the overall response displays the existence of a piezoelectric  response in this material.  The manifestation of the piezoelectric response may be a result of the relative closeness in energy of the anti-ferroelectric ground state of PbZrO$_3$.\cite{Singh}

For the (100)-PZT ceramic, we report an atomic distortion along the (100) direction.  As stated in Table II, the Ti atom is positioned $\sim$0.22\AA$\:$ above the plane containing the equatorial O atoms.  Also the Zr atom lies $\sim$0.45\AA$\:$ above its equatorial O atom plane.  This means that before application of the tetragonal strain, the upper halves of the MO$_6$ octahedra are compressed and their lower halves expanded.   Upon application of compressive uniaxial strain, we again find both expansion of the halves of the octahedra containing metal atoms and compression of the empty halves.  This results in metallic and oxygen motion opposing the ferroelectric distortion.  These atomic motions are very similar in nature to the reported responses in the tetragonal PbTiO$_3$.  Once again we note that the application of the uniaxial compression is done in a manner that will lead to a structural transformation towards a ceramic comprised of cubic-like sub-units.  A deformation that favors a para-electric phase will decrease the ferroelectric effect.  Therefore we conclude that application of a uniaxial tetragonal compression causes a piezoelectric response in (100)-PZT.  It is important to note that the structure of our (100)-PZT maintains the 4-fold rotation axis, a feature that would not allow any atomic motions along (111).  We may find composite atomic motions in the zirconate sub-unit, similar to the motions predicted in the PbZrO$_3$ crystal, upon removal of this symmetry constraint.  

The atomic distortions for the rhombohedral (111)-PZT equilibrium structure lie along the (111) direction.  In the undeformed superlattice, we find significant Ti and Zr displacements along the (111) relative to the Pb atoms.  Upon uniaxial compression we do not find any motion of the Ti or Zr atoms.  Instead we find complex Pb and O motions.  We report alternating motions of Pb atoms primarily along (100) as well as O motions along all three lattice directions.  The origin of O motion in (111)-PZT deserves particular attention.  The motion along (100) and ($\overline{1}$00) of the oxygen atoms that lie directly above and below the Zr and Ti atoms can be predicted from either the dilation or tetragonal responses.  However, the complex motion of the other oxygen atoms can only be ascertained from a simultaneous analysis of both responses.  As an illustration, we examine the motion of the O atom located near ($\frac{1}{4}$,0,$\frac{1}{4}$).  Because of its position relative to the expansion region depicted in the dilation response, it has a component of atomic motion along the (001) direction.  However, according to the tetragonal response, the same atom also moves along the (0$\overline{1}$0) direction due to the pairing of an oblate response centered in the TiO$_6$ octahedron and a prolate response centered in the adjacent ZrO$_6$ octahedron.  This same type of analysis can be done for all the oxygen atoms lying equatorial to the Zr and Ti atoms and can be used to predict their atomic motions due to uniaxial compression.  Based upon these atomic motions, it is possible to characterize the piezoelectric response in (111)-PZT.  The response can be summarized as complex motions of the Pb and O sub-lattices against the almost stationary Zr/Ti atomic positions, and not a simple move toward a higher symmetry structure, as seen in the previously described crystals.\vspace {5 mm}
\centerline{\bf V.  CONCLUSIONS}\vspace {3 mm}

In this density-functional study, we have examined the induced local stress fields resulting from an externally applied homogeneous uniaxial deformation of two distorted perovskite structures, PbTiO$_3$ and PbZrO$_3$.  We have found in the case of tetragonal PbTiO$_3$ the existence of a piezoelectric response involving atomic motions of the Ti and axial O atoms upon application of a tetragonal compressive strain.  We also report a piezoelectric response in the rhombohedral PbZrO$_3$ due to atomic motions along the (111) atomic distortion and (100) uniaxial strain directions.  In addition we have investigated the stress fields of the (100)-PZT and (111)-PZT superlattices.  We find that the stress responses of the (100) superlattice display similar features to those of the simpler distorted perovskites, most notably in the motion of the Ti and Zr atoms.  However in the (111)-PZT crystal, we have identified features not represented in any other crystal studied.  We find complex Pb and O motions against a fixed Zr/Ti sub-lattice upon application of a tetragonal stress.  Application of stress along a direction not parallel to the superlattice ordering and ferroelectric distortion directions gives rise to a complicated stress-field pattern.

This study demonstrates the utility of stress fields and their applicability to ferroelectric phenomena.  The construction of the local stress field is computationally tractable and provides an intuitive way to visualize and understand the response of a crystal to applied stress.  Furthermore, we feel that stress-field analyses done in conjunction with spontaneous polarization studies will broaden the understanding of piezoelectric materials.

\vspace{3 mm}
\centerline{\bf ACKNOWLEDGMENTS}\vspace{1 mm}
We would like to thank Karin Rabe for illuminating and valuable discussions on various aspects of the work.  In addition we would like to thank Steven P. Lewis, Eric J. Walter, Lewis D. Book, and Mark Feldstein of the University of Pennsylvania for their help in the preparation of this work and manuscript.  We also would like to thank John Shalf of the National Center for Supercomputing Applications for his help with the visualization of the stress fields using AVS 5.02.\cite{AVS}  This work was supported by the Laboratory for Research on the Structure of Matter and the Research Foundation at the University of Pennsylvania as well as NSF grant DMR 93-13047 and the Petroleum Research Fund of the American Chemical Society.  Computational support was provided by the National Center for Supercomputing Applications and the San Diego Supercomputer Center.

\centerline{\underline{FIGURES}}\vspace{0.10 in}
\noindent FIGURE IA. Dilation stress-field response to a uniaxial tetragonal deformation of tetragonal PbTiO$_3$.  (blue=lead, red=titanium, green=oxygen).

\vspace{0.25 in}
\noindent FIGURE IB. Tetragonal stress-field response to a uniaxial tetragonal deformation of tetragonal PbTiO$_3$.  (blue=lead, red=titanium, green=oxygen).

\vspace{0.25 in}
\noindent FIGURE IIA. Dilation stress-field response to a uniaxial tetragonal deformation of rhombohedral PbZrO$_3$.  (blue=lead, yellow=zirconium, green=oxygen).

\vspace{0.25 in}
\noindent FIGURE IIB. Tetragonal stress-field response to a uniaxial tetragonal deformation of rhombohedral PbZrO$_3$.  (blue=lead, yellow=zirconium, green=oxygen).

\vspace{0.25 in}
\noindent FIGURE IIIA. Dilation stress-field response to a uniaxial tetragonal deformation of (100)-Pb(Zr$_{0.5}$Ti$_{0.5}$)O$_3$.  (blue=lead, yellow=zirconium, red=titanium, green= \ oxygen).

\vspace{0.25 in}
\noindent FIGURE IIIB. Tetragonal stress-field response to a uniaxial tetragonal deformation of (100)-Pb(Zr$_{0.5}$Ti$_{0.5}$)O$_3$.
  
\noindent (blue=lead, yellow=zirconium, red=titanium, green= \ oxygen).

\vspace{0.25 in}
\noindent FIGURE IVA. Dilation stress-field response to a uniaxial tetragonal deformation of (111)-Pb(Zr$_{0.5}$Ti$_{0.5}$)O$_3$.  (blue=lead, yellow=zirconium, red=titanium, green= \ oxygen).

\vspace{0.25 in}
\noindent FIGURE IVB. Tetragonal stress-field response to a uniaxial tetragonal deformation of (111)-Pb(Zr$_{0.5}$Ti$_{0.5}$)O$_3$.
 
\noindent (blue=lead, yellow=zirconium, red=titanium, green= \ oxygen).

\end{document}